\documentstyle[twocolumn,aps,epsfig]{revtex}
\begin{document}
\narrowtext
\noindent
{\bf{Comment on ``Spin Dynamics of the 2D Spin $\frac{1}{2}$ Quantum 
Antiferromagnet
Copper Deuteroformate Tetradeuterate (CFTD)'' :}} 

In a recent Letter \cite{Ronnow01} R{\o}nnow {\it{et al.}} presented 
a neutron scattering study of the magnetic excitations of 
the antiferromagnet 
Cu(DCOO)$_2$ $\cdot$ 4D$_2$O for temperatures up to $T \approx J/2$,
where $ J \approx 73 K$ is the nearest neighbor exchange energy.
Due to the smallness of the interlayer-coupling and 
anisotropy-energies, this system is an almost ideal
realization of a  square lattice nearest neighbor
spin $S = \frac{1}{2}$ quantum Heisenberg antiferromagnet.
R{\o}nnow {\it{et al.}} paid particular attention to the temperature-dependence
of the dispersion and the damping of 
magnons with wave-vectors ${\bf{k}}$ close to the boundary of the 
magnetic Brillouin zone.
They observed a ``surprisingly good'' 
agreement of the data  for the temperature-dependence
of the magnon-damping rate $\Gamma$
with the formula
 \begin{equation}
 \Gamma_{\rm long}  ( T ) \approx
 {v_s ( T )}/{  \xi ( T )}
 \; ,
 \label{eq:gammalong}
 \end{equation}
which expresses the damping of {\it{long-wavelength}} magnons 
in terms of the spin-wave velocity $v_s ( T )$ and the antiferromagnetic 
correlation length $\xi ( T )$. 

In this Comment we point out that 
the agreement of the experimental data with Eq.\ (\ref{eq:gammalong})
is  accidental, because a comparison with
Eq.\ (\ref{eq:gammalong}) is only meaningful for {\it{long-wavelength}} 
magnons \cite{Tyc90}, whereas
the experimental data
give  the damping $\Gamma_{\rm short} ( T )$ of 
{\it{short-wavelength}} magnons, 
with wave-vectors close to the
boundary of the magnetic Brillouin zone.
The experimentally relevant  damping rate of
short-wavelength magnons for 
spin $S$ Heisenberg antiferromagnets has been calculated in Ref. \cite{Kopietz90}.  For 
wave-vectors in the regime $ | {\bf{k}} | \gtrsim   ( T a / v_s )^{1/3} ( 2 \pi / a ) $
(which includes the zone boundary)
the damping is \cite{Kopietz90,footnote1}
 \begin{equation}
 \frac{ \Gamma_{\rm short}   ( T ) }{J } = 
 \frac{ \pi}{ 24 } \frac{ Z ( | {\bf{v}}_{\bf{k}} | ) }{ S^4 }
 \left( \frac{ T }{ J } \right)^3 \left[ 1 + O \left( \frac{1}{2S} , 
 \frac{T}{ 4 S  J } \right) \right]
 \; ,
 \label{eq:damp}
 \end{equation}
where the numerical factor $ Z ( | {\bf{v}}_{\bf{k}} | )$ depends  on the length
of the dimensionless magnon-velocity
$  {\bf{v}}_{\bf{k}} 
 = \sqrt{2} \nabla_{\bf{k}} \epsilon_{\bf{k}}$ \cite{footnote2}.
From the numerical evaluation of $ Z ( | {\bf{v}}_{\bf{k}} | )$ \cite{Kopietz90}
it is easy to see that  $Z  ( | {\bf{v}}_{\bf{k}} | ) \approx Z ( 0 ) = 1$
for magnons with wave-vectors 
in the vicinity of the zone boundary, where $| {\bf{v}}_{\bf{k}} | \ll 1$.
Note that Eq.\ (\ref{eq:damp}) contains no adjustable parameters.

\begin{figure}
\begin{center}
\epsfig{file=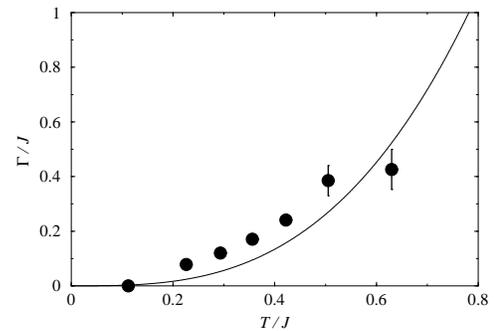,width=45mm,angle=270}
\end{center}
\caption{
Solid line: 
temperature-dependence of the damping of the zone-boundary magnons, see
 Eq. (\ref{eq:damp}) for $S = 1/2$.
The circles are the experimental data from Fig. 3 of Ref. [1].
}
\label{fig:damp}
\end{figure}

Taking into account that 
higher orders in $1/S$ and in $T / (SJ)$ have been neglected in Eq.\ (\ref{eq:damp}),
the agreement with the experimental data \cite{Ronnow01} 
is excellent, see Fig.\ \ref{fig:damp}.   
Note that Eq.\ (\ref{eq:damp}) is  based on the evaluation of
Fermi's golden rule, using for the relevant matrix elements 
the bare vertices describing magnon-magnon scattering within the
Dyson-Maleev transformation \cite{Kopietz90}. 
The fact that the resulting damping rate is in
quantitative agreement with the experiment 
indicates that  the effective 
interaction between antiferromagnetic magnons 
at short wavelengths is weak and hence can be 
treated perturbatively, without taking vertex corrections into account \cite{Kopietz97}. 
On the other hand, at long wavelengths 
the effective magnon-magnon interaction becomes strong
and prohibits the propagation  of stable magnons
over length scales exceeding the correlation length $\xi ( T )$.
This is the physics behind the  estimate (\ref{eq:gammalong}) for the
magnon damping rate at long wavelengths, which has nothing to do with the
short-wavelength damping due to weak magnon-magnon scattering
given in Eq.\ (\ref{eq:damp}).


This work was supported by the DFG via Forschergruppe FOR 412, Project No. KO 
1442/5-1.

%

\vspace{3mm}
\begin{center}
Peter Kopietz and Ivan Spremo\\
Institut f\"{u}r Theoretische Physik, Universit\"{a}t Frankfurt, \\
Robert-Mayer-Strasse 8, 60054 Frankfurt, Germany\\
\mbox{} \\
\end{center}
\noindent
Received: August 16, 2001\\
PACS numbers:  75.10.Jm, 05.70.Jk, 75.50.Ee
%
%
%

\begin{thebibliography}{99}
\bibitem{Tyc90}
More precisely, Eq.  (\ref{eq:gammalong}) can only be correct for
$   | {\bf{k}} | \ll 2 \pi / \xi$. 
Magnon damping in the regime $ 2 \pi / \xi \lesssim  | {\bf{k}} | \ll  
2 \pi / a$ (where $a$ is the lattice spacing)
has been discussed by
S. Ty\v{c} and B. I. Halperin, Phys. Rev. B {\bf{42}}, 2096 (1990).
\end{thebibliography}
\end{document}